\documentclass[runningheads]{llncs}

\usepackage[numbers]{natbib}
\usepackage{graphicx} 
\usepackage{float}
\usepackage{algorithm}  
\usepackage{algpseudocode}
\usepackage{color}
\usepackage[dvipsnames,svgnames]{xcolor}
\usepackage{mathrsfs}
\usepackage{amssymb}
\usepackage[dvipsnames]{xcolor}
\usepackage{fontawesome}
\usepackage{wrapfig}
\usepackage{fancyhdr}
\usepackage{amsmath}

\usepackage{setspace}

\begin{document}

\title {History Matching under Uncertainty of Geological Scenarios with Implicit Geological Realism Control with Generative Deep Learning and Graph Convolutions}

\author{Gleb Shishaev\inst{1}\and Vasily Demyanov\inst{2}\and Daniel Arnold\inst{2}}

\institute{Tomsk Polytechnic University\\
\email{glebshishaev@gmail.com} \and Heriot-Watt University\\
\email{V.Demyanov@hw.ac.uk}}

\maketitle 

\begin{abstract}
\textit{The graph-based variational autoencoder represents an architecture that can handle the uncertainty of different geological scenarios, such as depositional or structural, through the concept of a lower-dimensional latent space. The main difference from recent studies is utilisation of a graph-based approach in reservoir modelling instead of the more traditional lattice-based deep learning methods. We provide a solution to implicitly control the geological realism through the latent variables of a generative model and Geodesic metrics. Our experiments of AHM with synthetic dataset that consists of 3D realisations of channelised geological representations with two distinct scenarios with one and two channels shows the viability of the approach. We offer in-depth analysis of the latent space using tools such as PCA, t-SNE, and TDA to illustrate its structure.}

\end{abstract}
 
\section{Introduction}
\label{intro}

Subsurface reservoir modelling is fundamentally a problem of uncertainty - how to describe, account for, and quantify. The geological properties are highly heterogeneous, with correlation patterns linked to the sedimentation, diagenesis, and tectonic history of the reservoir. Understanding this history enables us to create a representative geological model using tools such as geostatistics. However, there is often considerable ambiguity in the specific geological concepts themselves, which is the actual cause of the uncertainty. As geological interpretations are often poorly constrained by limited well data and indirect geophysical observations, geologists typically describe uncertainty through a range of different geological concepts which must be modelled separately. 

Traditionally, geostatistical methods are used to address spatial uncertainty in geological models. These methods depend on stochastic statistical simulations, which parameterise spatial correlation structures in various ways. These models are calibrated to static well data where possible and are directed by knowledge of the geological concepts. However, geostatistical models are inherently limited by their parametric nature. This complicates the exploration of model uncertainty across fundamentally different geological scenarios, as the parameter spaces of these models cannot be meaningfully sampled or explored. Likewise, the interpolation of the model responses between these different concepts is effectively impossible, as the parameter spaces may have different vectors and different dimensionalities. Furthermore, geostatistical models are frequently bound to lattice grid-based representations that may not correspond to the stratigraphic or curvilinear geometries typical in natural geological systems.

In addition to these structural limitations, geostatistical modelling often fails to capture the complex, nonlinear relationships between geological input parameters and the resulting flow responses. This becomes increasingly problematic in highly heterogeneous reservoirs, where non-stationarity and multi-scale features dominate system behaviour. Integrating a range of dynamic and static data in a meaningful way is therefore very challenging. As a result, there is growing interest in alternative modelling strategies that can more flexibly represent geological uncertainty and accommodate multiple competing conceptual scenarios.

Generative learning models offer a promising solution. These methods, particularly auto-encoders, learn to represent complex data distributions in a low-dimensional latent space. Once trained, an autoencoder can generate diverse, realistic model realisations by sampling from this latent space, without requiring an explicit, parameterised description of spatial structure. This enables a shift from defining model uncertainty through hard-coded geostatistical parameters to learning it directly from data that can represent different plausible geological concepts. Crucially, these models can be trained on ensembles that include geological realisations generated under entirely different modelling assumptions. Then calibrated thereafter to dynamic measurements (production data, well tests, tracer tests, etc.) with the search in the latent space.

This paper proposes a Graph-based Wasserstein Autoencoder (GWAE) architecture tailored to geological modelling and history matching. Using graph representations, GWAE preserves the topological and stratigraphic integrity of geological features, allowing the model to capture the intrinsic structure of curvilinear and discontinuous geological systems more naturally than grid-based methods. The generated models can still be grid-based, but there is no need to capture the grid geometry explicitly. The Wasserstein formulation further encourages the latent space to be smooth and continuous, facilitating efficient sampling and interpolation.

The key innovation of our approach is its ability to combine multiple geological scenarios, with different dimensionalities, parameters, and assumptions, into a single learnt ensemble. This enables the definition and exploration of geological uncertainty within and across a model/concept class. In doing so, GWAE provides a unified and probabilistically coherent feature space for geological history matching, overcoming the limitations of traditional geostatistics and opening new avenues for data integration and decision-making under uncertainty.

The application of generative models to geological modelling has gained increasing attention over the past decade. Early efforts primarily focused on Generative Adversarial Networks (GANs), which learn to produce realistic geological patterns by pitting a generator against a discriminator in a two-network architecture. Notable contributions include the work of \cite{Chan_2019, chan2019parametrization, chan2018exemplarbased, dupont2018generating, mosser2019deepflow}, who demonstrated that GANs can generate complex geological models and reconstruct fine-scale porous media structures. Further developments by \cite{Laloy2018, LALOY2017387}  and \cite{SunChao2023} showed that GANs can also reproduce intricate depositional features such as meandering fluvial systems and reduce the dimensionality of inverse problems using specialised variants like spatial GANs (SGANs). While GANs often produce high-quality outputs, they are prone to issues like mode collapse, which limits the diversity of generated models and can lead to underrepresentation of geological uncertainty and/or unrealistic outputs.

Variational Autoencoders (VAEs) present a structured and probabilistic method for generative modelling. They convert data into a latent distribution, usually a multivariate Gaussian, allowing for the generation of new samples that can be decoded into realistic geological realisations. This creates a continuous and interpretable latent space, enabling smooth interpolation between different geological scenarios and facilitates efficient sampling. Research by \cite{Arauco2017, CANCHUMUNI201987, Arauco2021, LEE20181007, Jiang2021} demonstrates that VAEs excel in tasks requiring reduced parameterisation and uncertainty quantification. Furthermore, comparative studies by \cite{Bao2022} reveal that although GANs are often superior in replicating geological structures with limited training data, VAEs perform better in data assimilation tasks, particularly when large datasets are used. Their latent spaces accurately represent the geological variability essential for history matching and forecasting.
 
In contrast to GAN, VAE creates a structured and continuous latent space (LS) by using a probabilistic Encoder that maps input data to a distribution, typically a multivariate Gaussian. This allows for smooth interpolation and generation of input data variations. VAEs can sample from the LS and decode these samples to generate new, coherent, and diverse data instances. The structured nature of the latent space enhances interpretability, as each dimension can represent specific features or variations of the data.

Despite their advantages, most VAE-based approaches in geological modelling rely on grid or lattice data representations, which can misrepresent the complex stratigraphic geometries and curvilinear features found in natural systems. To overcome this issue, we propose a graph-based VAE architecture that represents geological models as graphs, where nodes and edges can encode various physical properties such as porosity, permeability, and facies type. Graph Convolutional Networks offer a powerful method for capturing spatial correlations and topological structures at multiple scales, allowing the model to learn from geological data more effectively than traditional lattice-based architectures. This enables a richer and more flexible representation of geological features in the latent space.

The generative model provides the means for geomodel calibration to production history without losing geological realism. 
The issue of implicitly controlling geological realism is a significant consideration when employing generative machine learning techniques, as these techniques do not possess inherent functionality that must be developed separately. The earlier contributions also addressed the problem of constraining the parameter space. 

The motivation for this work lies in developing a generative modelling framework that maintains geological realism while enabling efficient uncertainty quantification and history matching. One of the core challenges in applying generative models to reservoir modelling is ensuring that generated realisations remain plausible and geologically consistent when conditioned on production data. Earlier studies \cite{Demyanov_2018, Jung_2013, mosser2019deepflow} have highlighted the difficulty of controlling the realism of generated outputs without explicitly encoding geological constraints. Our approach addresses this by leveraging the learnt structure of the latent space, which captures geologically meaningful variations and correlations across scenarios. This space can be interrogated and constrained to produce plausible realisations during data assimilation. 

The objectives of this work are therefore threefold:
\begin{itemize}
    \item Using generative machine learning with graph convolutional networks to represent diverse geological models within a unified parametric space and to implicitly capture their dependencies;
    \item Accounting for uncertainty across different geological scenarios during history matching;
    \item Developing tools that ensure geological realism using the latent variables of a generative model.
\end{itemize}

\section{Methodology}
In this section, a methodology for generating reservoir models conditioned on production and static well data is presented using graph-based VAE with control of geological realism via LS. We will also present tools for analysing the internal structure of the latent space, such as PCA, t-SNE, and TDA, as this is required for the effective performance of the generative model.

Using VAE for history matching involves the following stages:

\textbf{Stage One: Re-parameterise the geological system by defining the minimum number of model parameters under prior uncertainty.}
This stage involves creating a prior training dataset that describes the entire space of uncertainty. The architecture of the generative model implies that the LS, that connects the \textit{Encoder} and \textit{Decoder}, determines the minimal set of model parameters (dimensions) that can describe the object under uncertainty and implicit dependencies between parameters (e.g. porosity-permeability).

\textbf{Stage Two: Forward modelling of a reservoir from the reduced parameters LS.}
After the neural network is trained and capable of reproducing the reservoir models of the training dataset and generating new models reliably, we can exclude the \textit{Encoder} from further consideration. New realisations are considered reliable if they are covered by various geological concepts incorporated in the prior ensemble that covers the uncertainty space. This constraint is imposed by the LS, which confines the newly generated instances to the space of models geologically consistent with the prior. 

The \textit{Decoder} essentially serves as a forward model, allowing us to translate the parameters into a physical representation of the object — a geological model. The difference from geostatistical methods is that the parameters of the latent space implicitly take into account the interdependencies between model parameters, thus limiting the variety of generated models to only realistic across different concepts.

\begin{figure}
    \centering
    \includegraphics[width=0.75\textwidth]{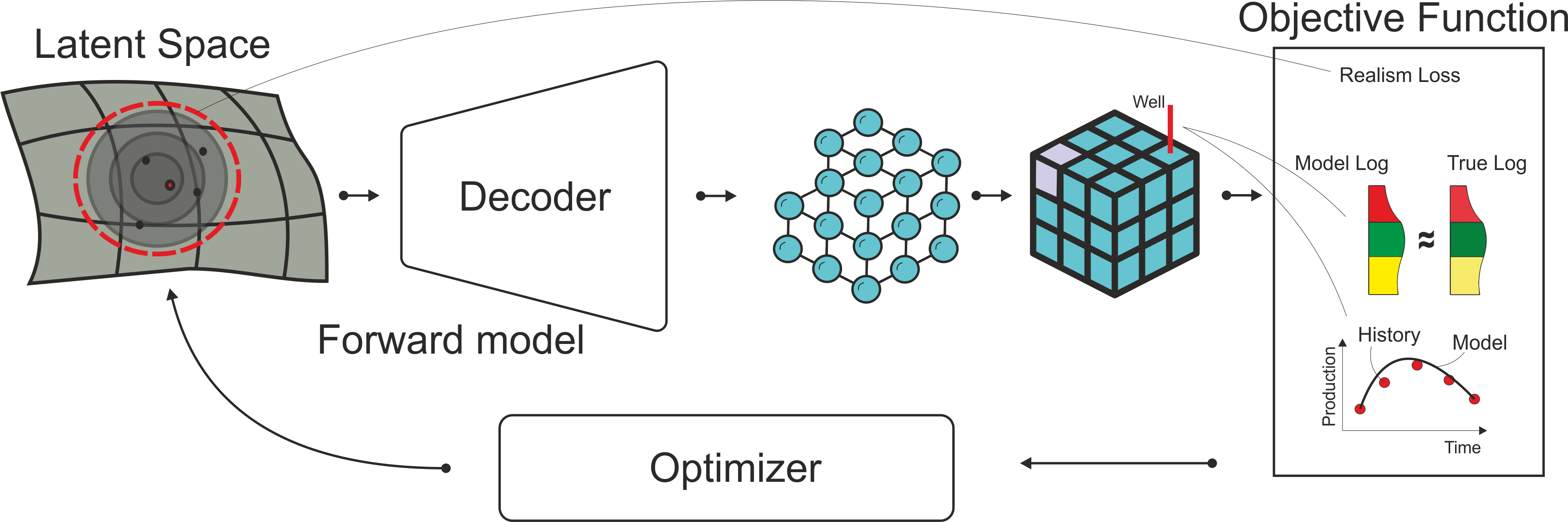}
    \caption{A latent vector is sampled from the LS, which is decoded to the geological model. Three-component OF then calculated and provides information for the optimiser about the next optimisation step in the LS}
    \label{fig:step three}
\end{figure}

\textbf{Stage Three: Inverse modelling to infer the model that matches both static and dynamic data.}
The final step involves searching for the location in the LS that matches the static and dynamic data after \textit{Decoder} reconstruction. Thus, it involves an optimisation process aimed at minimising the objective function (OF), Figure \ref{fig:step three}. 

Since the optimisation process through the latent space does not involve an built-in control of geological realism, such control can be imposed through the objective function being minimised (see Section 2.4). 

This approach of HM through a LS provides the following opportunities:
\begin{itemize}
    \item Provide common parameterisation across different scenarious, while also capturing complex, non-linear relationships between input parameters; 
    \item Effective model update with the reduced dimensionality of the model description in the LS with geological realism control;
\end{itemize}

\subsection{Importance of Geometric Deep Learning}

VAE's \textit{Encoder} and \textit{Decoder} by default have a 'classical' lattice-based CNN architecture. CNNs were initially developed for regularly structured (Euclidean) data, such as images, texts, etc., which should be perceived as a limitation. Geological structures often exhibit intricate spatial relationships, connectivity, and dependencies that graph representations can effectively capture, unlike Euclidean data types. When depicting a geological object using a graph format, employing grid cells is unnecessary, particularly for inactive cells whose sole purpose is to establish a regular grid. For intricate geological forms such as jigsaw or labyrinth types, or those with complex structural characteristics, there can be a substantial quantity of these cells that provide no valuable information.

Graph-based convolutions are highly effective for geological modeling due to their capacity to preserve the topology of geological structures. In this approach, geological features are represented as nodes and their relationships as edges, forming a graph. This method aligns well with the intricate and interconnected nature of geological formations, enabling a more natural representation of geological topology. Consequently, it allows for precise and realistic modeling of subsurface processes, fault networks, and other complex geological phenomena.

\subsection{Graph Wasserstein Autoencoder (GWAE)}
The key feature of the proposed convolution operation on graph data types is that it occurs on sub-graphs of different dimensions, not just nodes \cite{morris2021weisfeiler}. Thus, it becomes possible to incorporate a hierarchical approach to graph processing, which allows taking into account and combining geological features at the scale of nodes and larger-scale objects in the form of sub-graphs. In combination with VAE's LS, this approach makes it possible to account for uncertainty more flexibly at different scales, ranging from geological scenarios to more localised ones. 

The hierarchy relates to MPS, which focusses on connected data and higher-order statistics. In geostatistics, techniques like Training Images/MPS use a lattice-based approach to define spatial patterns through state transitions on a grid. In contrast, the graph-based method emphasises data topology, capturing connections between nodes and sub-graphs. This enables insights into first- and second-order connections. Unlike MPS, which gathers statistics and infers distributions, the graph model incorporates a learning mechanism that adapts to complex relationships in the data, offering a dynamic framework for identifying higher-order statistical patterns in interconnected datasets through LS.

To calculate the properties of the target node $t$ for a layer $l+1$, all neighbouring nodes should be considered. The information from the neighbouring nodes is passed to a transformation operation ($Q$), then aggregated ($\sum$($\cdot$)) and transformed ($W$) regarding the target node $t$. This sequence of actions is performed for each node in the graph. Thus, matrices $Q$ and $W$ are the training parameters of the neural network.
\begin{align}\label{Spatial_filter_eq}
    &h_t^{(l+1)}=f^{l}(h_t^{(l)},h_n^{(l)})=\sigma(W^{(l)}h_t^{(l)}+Q^{(l)}h_n^{(l)})\\
    &h_n^{(l)}=\frac{1}{(n-1)}\sum_{n\ne t}h_i^{(l)}
\end{align}

To introduce the concept of graph-based hierarchy, it is necessary to modify equation \ref{Spatial_filter_eq} to include the neighbourhood of nodes and sub-graphs of different levels. Introduce the $k$-element subset $[V^k] \subset V$, where $s=\{s_1,...,s_k\}$ - $k$-set in $[V^k]$. So, the neighbourhood of $s$ is: $N(s)=\{t \in [V^k] | |s\cap t|=k-1\}$.
The local neighbourhood $N_L(s)$ consists of $t \in N(s)$, which has edges $(u, v) \in E$ for every $u \in s \setminus t$ and $v \in t \setminus s$. On the other hand, the global neighbourhood $N_G(s)$ is defined as $N(s) \setminus N_L(s)$.

Equation \ref{Spatial_filter_eq} can be defined as follows:
\begin{align}\label{My_graph_conv}
    h_t^{(l+1)}(s)=\sigma\big(W^{(l)}h_t^{(l)}(s)+\sum_{u\in N_L(s)}Q_L^{(l)}h_n^{(l)}(u)+\sum_{v\in N_G(s)}Q_G^{(l)}h_n^{(l)}(v)\big) 
\end{align}
Using matrices $W, Q_L, Q_G$ makes it possible to train a neural network taking into account local and global neighbourhoods. Additionally, using subsets of nodes $(s)$ as a single element allows hierarchy in training.

As a generative architecture we used regularised Autoencoders based on optimal transport theory, which is called Wasserstein Auto Encoders (WAE) and was introduced by \cite{WAE}. Similarly to conventional VAE, the objective of WAE is composed of two terms: the reconstruction cost and the regulariser.

The WAE objective function compared to VAE.
\begin{gather}
    D_{VAE}\big(P(x),P(X|Z)\big) = \inf \mathbb{E}_{P_X}\big[KL\big(Q(Z|X)||P(Z)\big) - \mathbb{E}[\log P(Z|X)] \big] \\
    D_{WAE}\big(P(X),P(X|Z)\big) = \inf \lambda D\big(Q(Z|X),P(Z)\big) + \mathbb{E}_{P_X}\mathbb{E}_{Q(Z|X)}\big[c\big(X,G(Z)\big) \big]
\end{gather}

GWAE consists of two neural networks: an \textit{Encoder} \big($Q(Z|X)$\big) and a \textit{Decoder} \big($P(X|Z)$\big), each of which is a sequence of graph convolutional networks. The training dataset, a set of unstructured graphs, is passed to the \textit{Encoder}, which performs convolutional operations. The result is the parameters of the distribution $Q(Z|X)$ which are the parameters of the LS. In the next step, a hidden vector $z$ is sampled from the distribution $Q(Z|X)$. If the GWAE is trained correctly, the vector $z$ contains enough information for the \textit{Decoder} to reconstruct $x$ correctly.

\subsection{Inner Geometry of a Latent Space}
\label{Riemann}

We assume that the geological multivariate spatial distribution of properties can be encoded into a low-dimensional representation, or LS. 
On the other hand, a LS should be viewed as a curved surface \citep{rubenstein2018latent}. Therefore, when analysing a LS, it is necessary to consider its curvature, which leads to errors using Euclidean metrics when working with latent representations. 

Replacing Euclidean "straight lines" with geodesics "shortest paths" for working with latent spaces GWAE will allow us to achieve the following goals: 
\begin{itemize}
    \item Implicit geological reliability control. Introduce a metric that corresponds to the density of the regions in the LS that will constrain the optimisation process to the high-density space, which means that the Decoder reproduces the most probable objects, which can be interpreted as the most realistic objects from a geological point of view, because such realisations prevail in the prior.
    \item Shortest paths. Allows the selection of latent vectors that correspond to similar geological objects more confidently. 
    \item Smooth interpolation through the LS. Provides the ability to smoothly transition from one geological realisation to another through the LS while navigating across the LS to minimise the OF.
\end{itemize}
To implement this concept, it is necessary to include the notion of Riemannian spaces in the analysis.

The works of \cite{tosi2014metrics, Lawrence2005ProbabilisticNP, arvanitidis2021latent, shao2017riemannian} were among the first to focus on the analysis of latent spaces generated by generative neural networks in terms of their internal non-linearity. They showed that taking into account internal non-linearity improves the quality of sampling latent vectors $z$ of the LS $\mathcal{Z}$, which leads to an increase in the quality of generated objects $x$.

The present section introduces a metric for assessing the curvature of the hidden space, which was adapted from the work of \citep{b0c148db029248e894c41f24a6051390}.

Our dataset $\{x\} \in \mathcal{X} = \mathbb{R}^n$. Similarly, let us assume that $\mathcal{X}$ "lives" in some non-linear hidden space $\mathcal{Z}$.

Denote the Riemannian space by $\mathcal{Z} = \mathbb{R}^m$, since the hidden space of GWAE should ideally be as close to it as possible. Therefore, to take into account the curvature of the Riemannian space, we need to introduce a metric tensor $\mathbf{Z}:\mathcal{Z}\to S_+^m$. The values of the metric tensor should be small in regions of $\mathcal{Z}$ confirmed by observations ${x}$, otherwise $\mathbf{Z}$ should be large, indicating that in such regions of $\mathcal{Z}$ there was no data from the training set ${x}$.

By definition, the \textit{Decoder} is a function of the form: $f(Z)= P(X|Z) = \mu_{\theta}+\sigma_{\theta}\odot \epsilon$, where $\epsilon \sim \mathcal{N}(0,I_n)$. According to \cite{b0c148db029248e894c41f24a6051390}, $f$ should generate a pull-back metric \cite{Khlevniuk} in the space $\mathcal{Z}$, taking into account its internal geometry \citep{arvanitidis2021latent}. 
\begin{gather}\label{density_estimator}
\mathbf{Z}_f(z) = \mathbb{J}_f(z)^\intercal \mathbb{J}_f(z) = \mathbb{J}_{\mu}(z)^\intercal \mathbb{J}_{\mu}(z)+\mathbb{J}_{\sigma}(z)^\intercal \mathbb{J}_{\sigma}(z).
\end{gather}

This metric $\mathbf{Z}$ has several desirable properties. The first term considers the curvature of the hidden space $\mathcal{Z}$, while the second term increases with uncertainty. Therefore, regions of the hidden space without training data should have high metric values $\mathbf{Z}$.

\subsection{Objective function in HM process}
\label{section:OF_in_HM}
The essential components of the AHM process are an optimisation algorithm and an OF, which estimates the agreement of a model under consideration with the data at hand. With the development of generative neural networks as a method of matching geological models, the OF must control the quality of matching based on production data and static indicators such as well logs. Furthermore, even a good match does not guarantee the geological realism of the final model, as the spatial distribution of properties may be disrupted \citep{Demyanov_2018}. Following \cite{mosser2019deepflow}, the OF should consist of three main components: the dynamic production data mismatch ($Loss_{flow}$), the static conditioning well data mismatch (e.g., well-log derived data ($Loss_{static}$)), and the control of geological realism of the model in general ($Loss_{realism}$), that is accounts for the geological consistency of the modelled property distribution in the inter-well space:
\begin{gather}\label{Loss_for_GWAE}
Loss = w_f Loss_{flow} + w_s Loss_{static} + w_rLoss_{realism}
\end{gather}
Where $w_f, w_r, w_r$ are weights to provide additional flexibility to the overall loss function. The components $Loss_{static}$ and $Loss_{flow}$ are represented by the MSE \citep{Yeremian}.

\subsubsection{Realism loss function}

The last component of the loss function, equation \ref{Loss_for_GWAE} ($Loss_{realism}$), accounts for the realism of the generated models. It evaluates how close the generated patterns are to those attributed to the assumed geological concept. The concept is represented by LS mapping of the prior ensemble used for training and is initially generated by the geologist.

The GWAE training set is assumed to consist only of geologically realistic scenarios, which means the neural network is trained to reproduce the underlying geology correctly. However, LS ($\mathcal{Z}$), which is the “source” of geological realisations, has a probabilistic structure. Therefore, if we sample from low-probability areas, the reconstructed geological realisations may contain disrupted geology. This means that the optimiser should receive information about the probability of occurrence of a particular realisation. We replaced the probability estimation with an estimate of the density of LS $\mathbf{Z}_f(z)$ from the equation \ref{density_estimator}.

Thus, as the third term, the loss function will receive an estimate of density $\mathbf{Z}_f(z)$. If this estimation starts to increase, it means that the optimiser is starting to sample from areas of the LS with low density, i.e., areas with few training samples, indicating a low probability of occurrence of such a realisation. As the estimation of density decreases, the probability of occurrence of such a geological realisation will also decrease.

The experiments carried out have shown that the static and dynamic components of the OF work coherently, which is logical since the well log data is one of the most critical indicators for subsequent production. Meanwhile, the third component operates more against it as it imposes constraints on the optimisation process, restricting the search exclusively to the zone of geologically realistic scenarios, trading off "realism" for "less exploration" of the LS.
\section{Results - Application}

We present the experimental results of AHM using GWAE on a synthetic dataset that consists of 3D realisations of channelised geological models with two distinct scenarios with one and two channels. We will analyse the generative capacity of the model, the structure of the latent space, and show why accounting for its intrinsic non-linearity is an important factor. In addition, an ablation study will be presented to show the importance of considering geological realism in the optimisation process.

\subsection {Case dataset construction}
The initial step in geological modelling with generative neural networks involves creating a set of prior subsurface models based on existing information. The prior ensemble is used to train the model and generate the corresponding data and prediction variables. 

\begin{figure}
    \centering
    \includegraphics[width=1\textwidth]{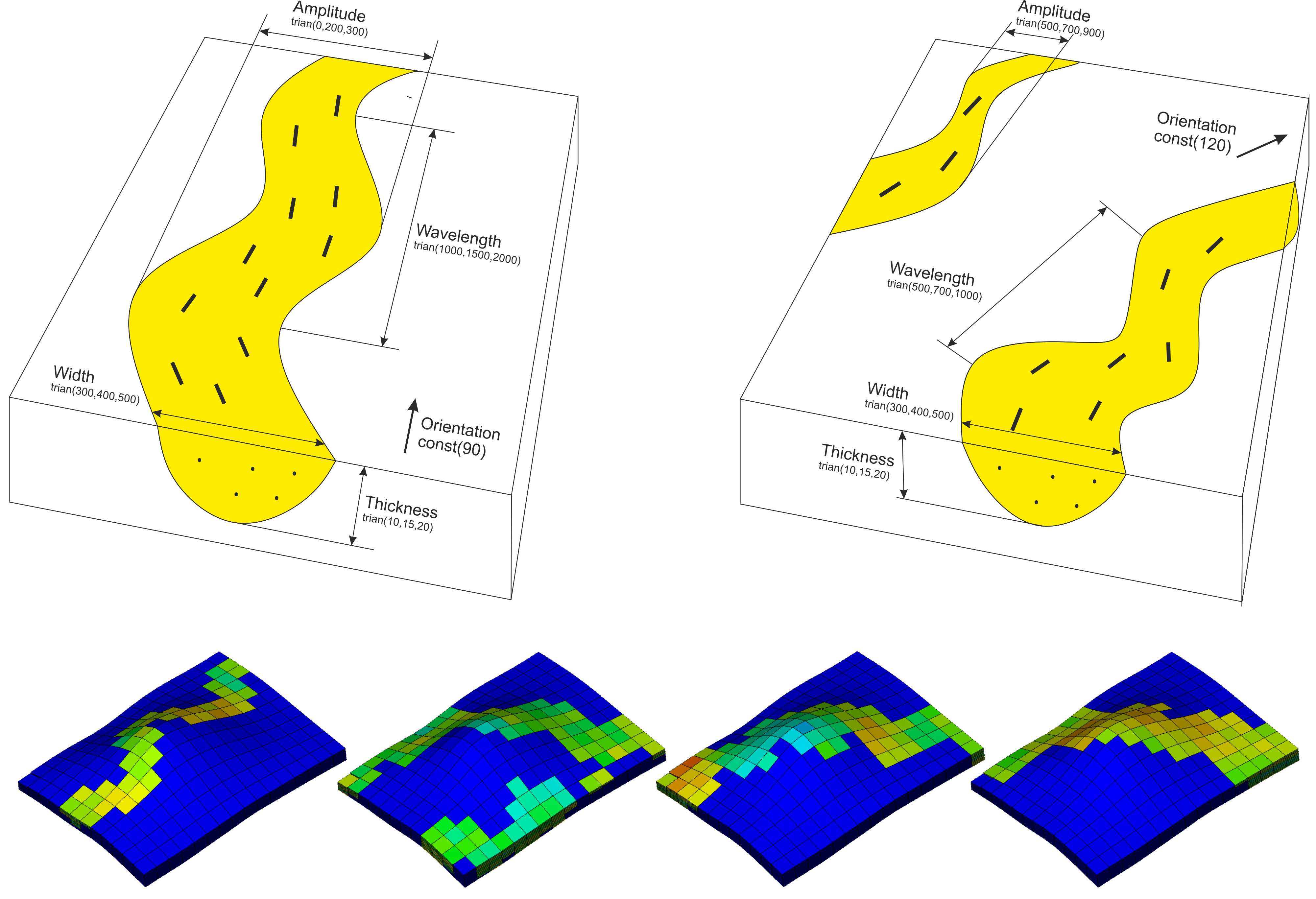}
    \caption{Settings of the channelised synthetic dataset. Two distinct geological scenarios: single and double channels are considered. The second row represents some examples from the dataset}
    \label{fig: Channelised case}
\end{figure}

The dataset consists of a set of three-dimensional models of porosity and permeability properties of a simplified channelised synthetic cubes, Figure \ref{fig: Channelised case}. A simplified fluvial setting was chosen as the base geological concept, allowing visual quality control of generated realisations. It consists of scenarios with one and two channels. The channel parameters for each scenario vary within different uncertainty ranges, Table \ref{table:channel_case}. The main uncertainty of this dataset is the uncertainty in the location of the channel bodies, their quantity, and channel parameters. Furthermore, the data set has uncertainty of the dependence between porosity and permeability. 

\begin{table}[htbp]
\caption{Table of parameters for dataset construction}
\label{table:channel_case}
\begin{center}
\begin{tabular}{ | c | c | c | } 
  \hline
  \textbf{Scenario}& \textbf{Single} & \textbf{Double} \\ 
  \hline
  Number of channels & one & two \\ 
  \hline
  Width, m & $[300; 500]$ & $[300; 500]$ \\ 
  \hline
  Thickness, m & $[10; 20]$ & $[10; 20]$ \\ 
  \hline
  Wavelength, m & $[1000; 2000]$ & $[500; 1000]$ \\ 
  \hline
  Amplitude, m & $[0; 300]$ & $[500; 900]$ \\ 
  \hline
  Orientation, deg & $90$ & $120$ \\ 
  \hline
\end{tabular}
\end{center}
\end{table}

Five thousand realisations were created, 2500 for each scenario with one or two channels. The dimension of the model is $x = 16$, $y=12$, $z=10$, which means that the dimension of the optimisation search space is $x\cdot y \cdot z \cdot NumberOfProperties = 3840$, since the optimisation algorithm needs to find values for porosity and permeability for each cell.

A properly trained GWAE should adhere to the following conditions: preserve the statistical distributions of porosity and permeability properties, maintain the petrophysical multivariate relationship between the modelled properties, and preserve the correct spatial geometry for the channel facies consistent with the prior regardless of the number of channels in any of the scenarios.

A flow model was prepared to carry out simulation experiments. The model consists of two phases: oil/water with a line drive scheme of one line of producers in the middle and two lines of injectors. The well performance calculation is set for 60 time steps. The main parameters of the flow simulation model are presented in Table \ref{table:Flow_simulation_parameters}.

\begin{table}[htbp]
\caption{Flow simulation parameters}
\label{table:Flow_simulation_parameters}
\begin{center}
\begin{tabular}{| c | c | c | } 
  \hline
  \textbf{Parameter}& \textbf{Value} & \textbf{Unit} \\ 
  \hline
  $P_{init}$ & $240$ & bar at OWC\\ 
  \hline
  $P_{prod}$ & $45$ & bar \\ 
  \hline
  $P_{inj}$ & $330$ & bar \\ 
  \hline
  $N_{wells}$ & $9$ & $-$ \\ 
  \hline
  OWC & $2460$ & $meter$ \\ 
  \hline
  Water viscosity ($\mu_w$) & $0.40$ & $cP$ \\ 
  \hline
  Oil viscosity ($\mu_o$) & $3.3$ & $cP$ \\ 
  \hline
  Water density ($\rho_w$) & $1020$ & $kg/m^3$ \\ 
  \hline
  Oil density ($\rho_o$) & $875$ & $kg/m^3$ \\ 
  \hline
  Connate Water Saturation & $-0.048 \cdot \ln(perm) + 0.5$ & $frac$ \\(SWL) & &\\
  \hline
  Critical Water Saturation & $1.1 \cdot$ SWL & $frac$ \\
  (SWCR) & &\\
  \hline
  Initial Water Saturations & $-0.18 \cdot \ln(J) + 0.57$ & $frac$ \\ 
  (SWATINIT) & &\\
  \hline
  Scaled Critical Oil-in-Water & if(SWATINIT=1, 0, 0.25) & $frac$ \\ 
  Saturations (SOWCR) & &\\
  \hline
  Maximum Water Saturation & if(SWATINIT=1, 1, 1-SOWCR) & $frac$ \\
  (SWU) & &\\
  \hline
\end{tabular}
\end{center}
\end{table}

\subsection{Generative quality of GWAE}

After training GWAE, we obtained a unified parameter space that includes variants of both geological scenarios. The \textit{Encoder} of GWAE decreased the initial dimensionality of 3840 to the dimensionality of a LS, which is 30. Several publications \citep{Automated_discovery, NIPS2004_74934548, guss2018characterizing} provide a dimensionality selection process, but this research went beyond the scope of the current work. If the dimensionality is reduced below 30, the performance of the \textit{Encoder} starts to deteriorate, indicating that the features become insufficiently informative for the correct generation of properties.
\begin{figure}
    \centering
    \includegraphics[width=1\textwidth]{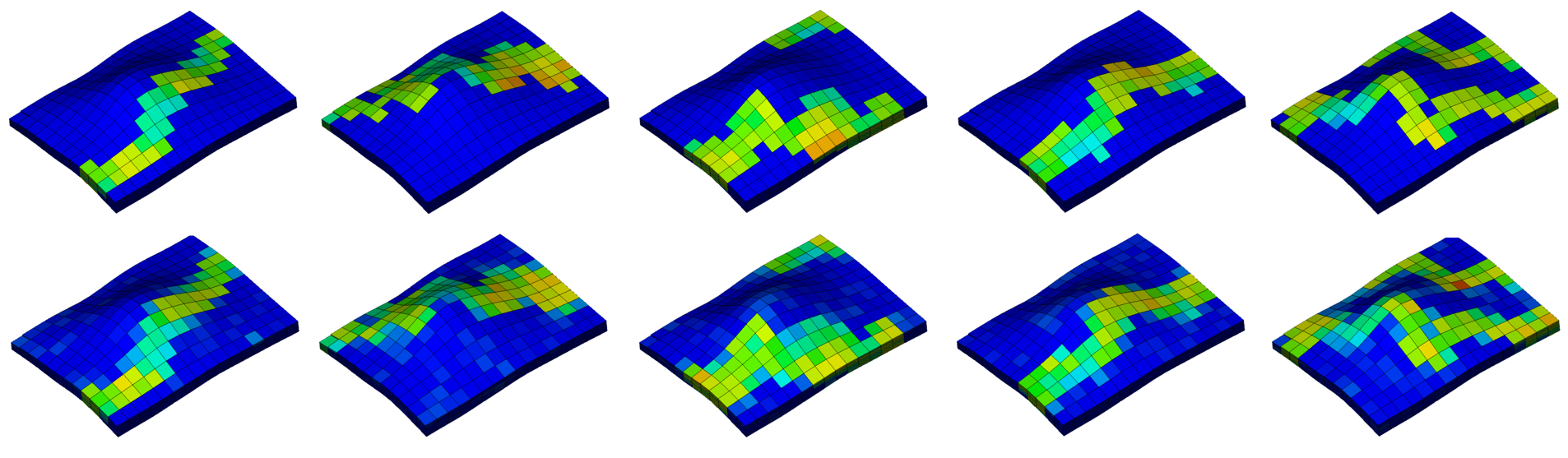}
    \caption{1: Examples of the initial porosity grid from the training dataset. 2: GWAE reconstructed. 3: GWAE preserves petrophysical relations and spatial propagation of properties under uncertainty of geological scenarios}
    \label{fig:fluvial_true_pred}
\end{figure}

The following Figure \ref{fig:fluvial_true_pred} shows random realisations of the training dataset and those reconstructed using the GWAE \textit{Decoder}. It can be concluded that the vectors of the LS are informative enough to reconstruct the spatial dependencies of the property distribution and petrophysical property relations.

\subsection{Latent Space inner structure}
The next stage in the exploration of the generative capabilities of GWAE is a more in-depth analysis of the LS. The following Figure \ref{fig:fluvial_LS} represents a visualisation of the 30-dimensional LS through dimensionality reduction methods, such as PCA \citep{Karl_1901, Hotelling1933AnalysisOA}, t-SNE \citep{t-SNE}, and TDA \citep{Edelsbrunner, Carlsson}. 

\begin{figure}
    \centering
    \includegraphics[width=1\textwidth]{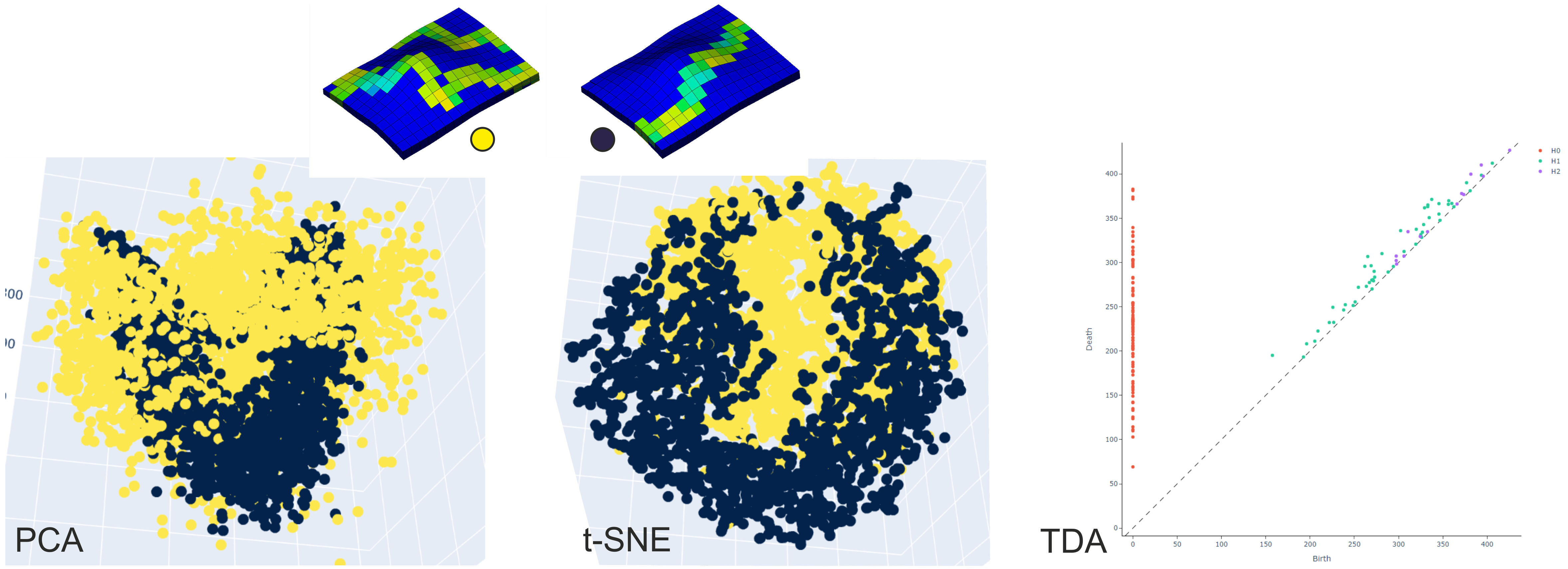}
    \caption{Visualisation of the 30-dimensional LS in the PCA,t-SNE, and TDA spaces. Each point in the PCA and t-SNE space represents a 30-dimensional vector characterising a geological realisation from the training dataset.
    Blue dots depict a geological scenario with one channel while Yellow dots represent scenarios with two channels. In the TDA space, the appearance and disappearance of connected components are displayed.}
    \label{fig:fluvial_LS}
\end{figure}

\subsubsection{PCA and t-SNE}
The space consists of a cloud of points comprising two large clusters. Each point represents a specific 30-dimensional coordinate of model location in the LS for each geological realisation from the training dataset, with the colour indicating a particular scenario. We can infer that if sampling is performed in a specific region of the LS, we can expect the generated geological realisations to belong to a certain scenario. 
\begin{figure}
    \centering
    \includegraphics[width=0.50\textwidth]{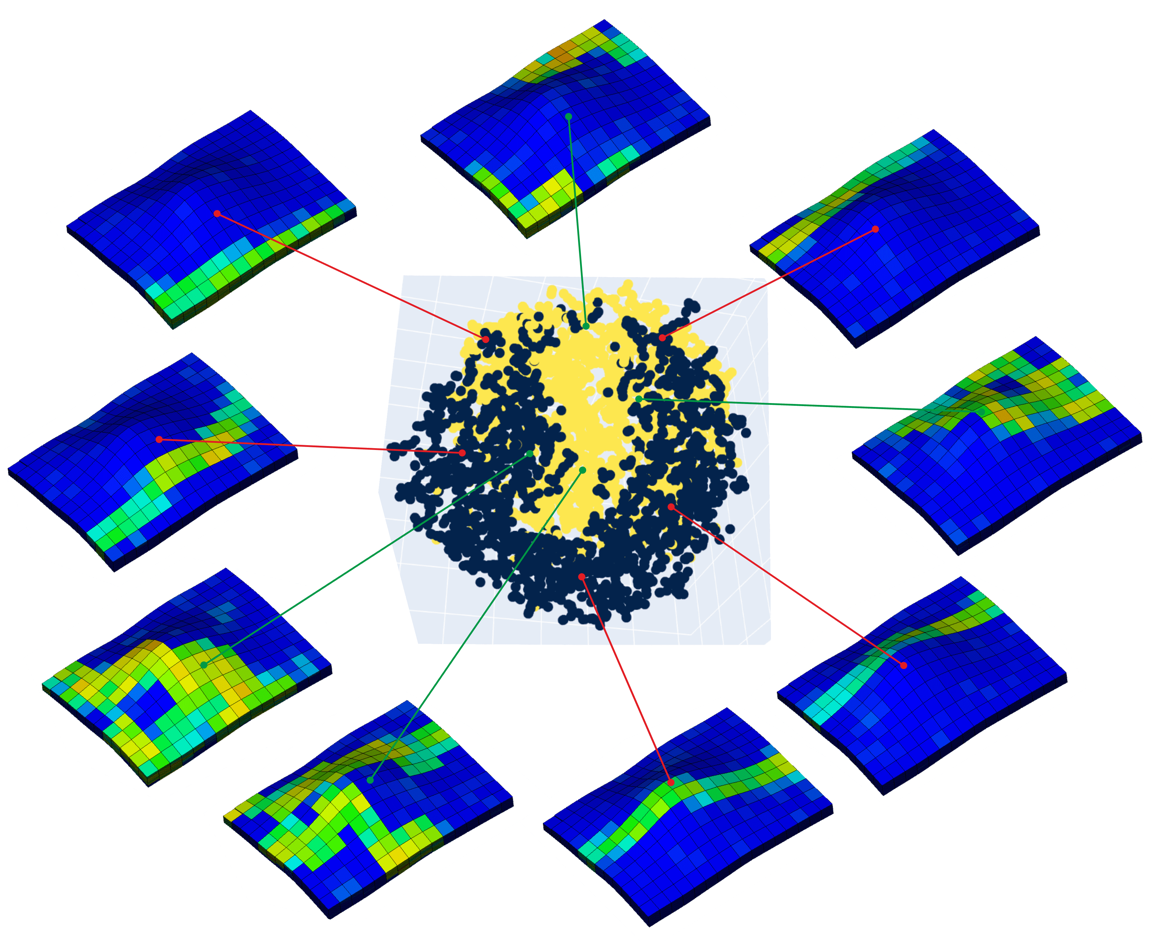}
    \caption{3-d t-SNE visualisation of the 30-d LS. Every point in this cloud is a sample from the training dataset. left: the LS in the scenario dimension, where red bars represent a scenario with one channel, and green bars represent two channels}
    \label{fig:flivial_manual_sampling_LS}
\end{figure}

Figure \ref{fig:flivial_manual_sampling_LS} shows geological realisations corresponding to specific regions of space. 
The LS possesses an internal structure reflected in the spatial arrangement of the channels. If we consider the U-shaped blue cloud of points, it can be observed that the channel 'shifts' from the southeast to the northwest. The yellow cloud of points exhibits a similar dependency. The northern part of the LS produces realisations where the channels are spread across opposite sides. In contrast, the central part of the LS allows for reconstructions with channel placements closer to the central area of the grids.

\subsubsection{TDA}
\begin{figure}
    \centering
    \includegraphics[width=0.5\textwidth]{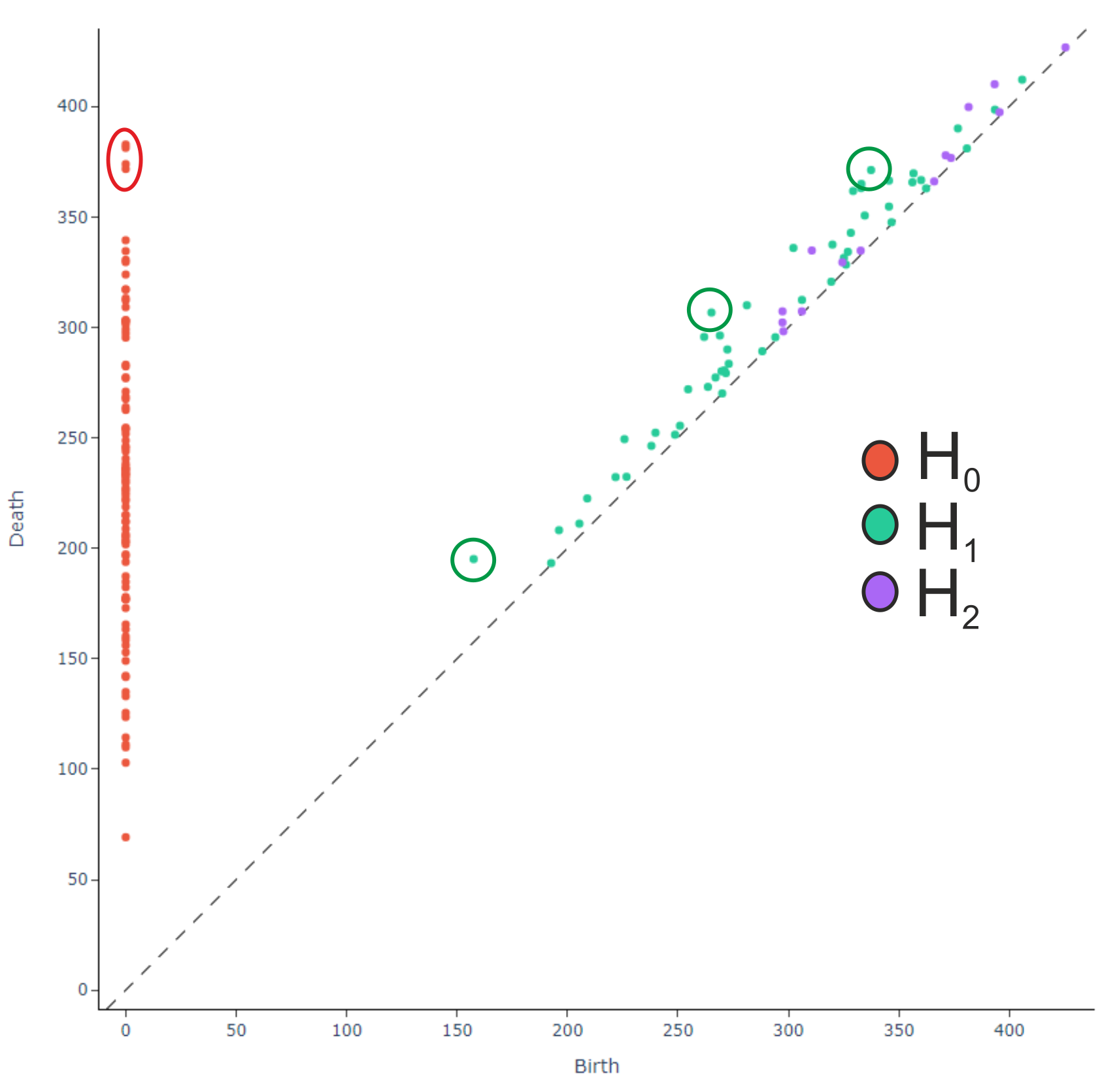}
    \caption{The LS in the context of TDA.
    Red circle denote points that are furthest from the rest on $H_0$, and green circles
    most stable rings ($H_1$). However, it cannot be concluded that these objects are global topological properties of the space.}
    \label{fig:flivial_TDA}
\end{figure}
Analysing the topology of the LS involves comprehending its geometry and structure without necessarily reducing its dimensionality. TDA enables the examination of complex, multidimensional, noisy data and the extraction of essential characteristics that offer insights into the shape of the object under consideration \citep{tierny2018topological}. The entire LS is analysed as a single object. To do this, refer to Figure \ref{fig:flivial_TDA}. It can be seen that in space $H_0$, the points are arranged along the vertical axis, indicating that the hidden space vectors are separated from each other by different distances, with some distinct objects marked by the red circle. 

However, in $H_1$, there are no pronounced objects, as the points are stretched along the diagonal more or less equally. Minor outliers are denoted by green circles, suggesting that there may be several rings in this dimension that do not represent global objects. In general, it can be concluded that the 30-dimensional LS lacks any distinct topological characteristics and is more akin to a simple cloud of points. In the context of TDA, one cannot conclude that geological scenarios are somehow distinguishable. Instead, both scenarios appear to form a single cloud.

\subsection{Variation of models along the shortest path trajectories in the LS}
\label{section:Fluvial_interpolation}
To assess the generative capabilities of the neural network, it is necessary to evaluate how well it can generate new objects that are not present in the training dataset. This assessment will use a variation of models along trajectories through the LS as an example. It is important to note that the LS is not linear inside. Therefore, the shortest path will not be defined by the Euclidean distance but rather by the Geodesic distance. 

\begin{figure}
    \centering
    \includegraphics[width=0.75\textwidth]{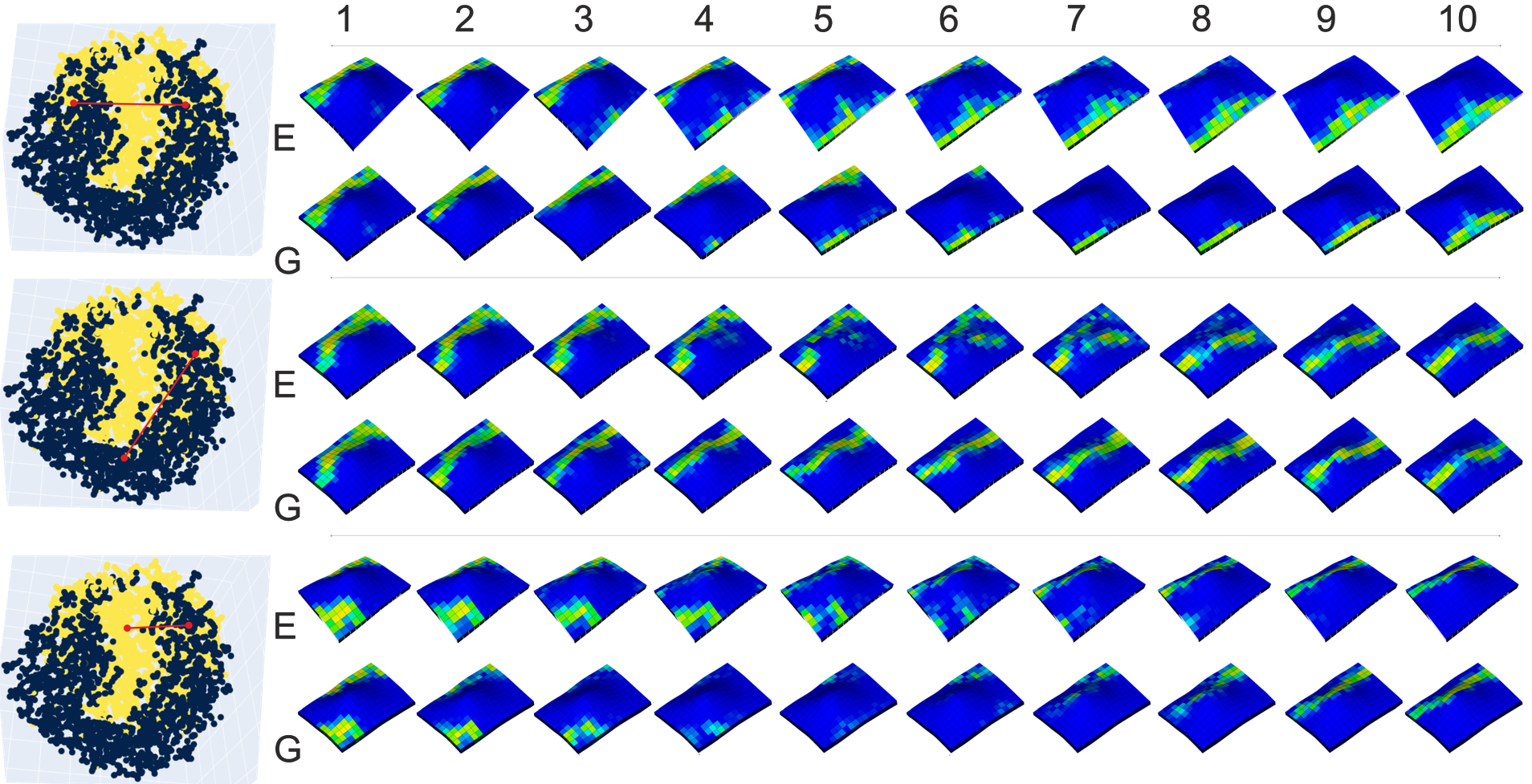}
    \caption{Variation of ensemble members along the shortest path trajectories in the LS. Left side: starting and final points in the LS. Right side: Euclidean (E) and Geodesic (G) interpolation options in 10 steps. It's noticeable that Geodesic interpolation enables the reconstruction of higher-quality geological realisations.}
    \label{fig:fluvial_interpolation}
\end{figure}

We selected two random geological realisations for the experiment and identified their positions in the LS. Then, we calculated the shortest paths through the LS using both Euclidean and Geodesic metrics. Ten realisations were chosen along these paths and reconstructed back into geological models using the \textit{Decoder}. Figure \ref{fig:fluvial_interpolation} shows three such experiments. It can be observed that Geodesic interpolation provides higher-quality realisations, while Euclidean interpolation becomes less confident at intermediate steps. 

In the Figure, special attention should be paid to the third case – the transition from one concept to another. Here, it can be seen that Euclidean interpolation practically deviates into unrealistic implementations from steps 4 to 8, while Geodesic interpolation only shows non-quality implementations at steps 7 and 8. This behaviour can be explained as follows: According to the TDA analysis, objects in the space are distant from each other in $H_0$. These objects can be geological scenarios. Therefore, when interpolating between scenarios, we inevitably enter a low-density space, resulting in unrealistic geological variations. However, the Geodesic metric attempts to 'stay' in dense regions of the LS as long as possible.

It should be emphasised that the intermediate realisations are not instances of the training data set. This implies that the quality of the GWAE \textit{Decoder's} performance depends on the quality of the training process and the hidden vector used for reconstruction. It cannot be expected that a generated object will be of high quality if the latent position was sampled from a region of the LS where no training examples are present. 

\subsection{History matching through the latent space}
The optimisation experiment was conducted through the LS conditioned on static and dynamic well data. Static conditioning data are "hard" only in a geostatistic sense. In reality, conditioning data derived from wireline log interpretation are subject to uncertainty that is rarely considered. Furthermore, well-log support volume does not make it "hard" conditioning data with the grid-cell support size. Therefore, static well data should be treated as "soft" approximate conditioning that is not traditionally followed in geostatistic reservoir modelling. GWAE provide "soft" conditioning to both static and dynamic data through minimisation of the OF. 

A reference model was chosen, which will be unknown except for well-log and production data. Then, an optimisation process was started by iteratively generating the population of models and comparing it with the reference well data, consequently minimising the OF.

The LS has various densities, so the density is considered to reflect the probability of a particular geological setting. We estimated the density of LS in different regions by the parameter metric $\mathbf{Z}$, which provides an estimate of the curvature of the LS and increases the uncertainty estimate as we move away from dense regions of the LS covered by training data. If the value of Realism loss is high, the optimisation algorithm sampled from the subspace is of low density, i.e., highly unlikely realisation or geology of low reliability. Otherwise, the sampling process is conducted in regions of high density, that is, the realisation is likely.

\begin{figure}
    \centering
    \includegraphics[width=0.75\textwidth]{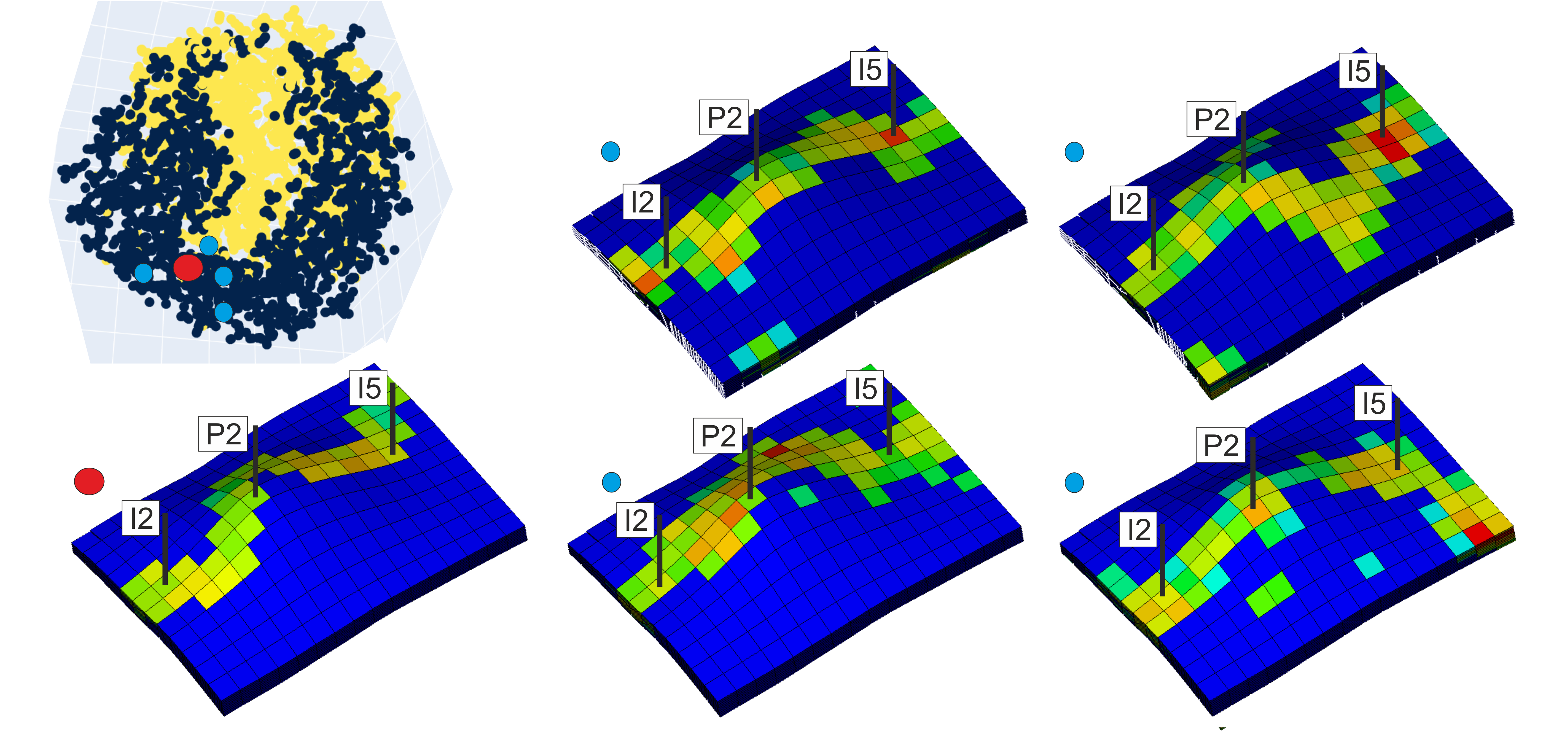}
    \caption{Visualisation of the reference porosity property 
    (red dot) and the cubes obtained through the optimisation process
    (blue dots). Due to high uncertainty in regions without well data, the geological realisations have not been completely similar to our reference model}
    \label{fig:fluvial_Geology_matched}
\end{figure}

The CMA-ES optimisation algorithm was chosen to search LS \citep{hansen2023cma}. It takes the results of each generation and adaptively increases or decreases the search space for the next generation. We launched four optimisation processes, each composed of 100 iterations, with a population size of 51 samples. The optimisation process has found various realisations that all meet the following requirements: similarity in static and dynamic wells data along with geological realism of every found solution. The results of the optimisation process are presented in Figure \ref{fig:fluvial_Geology_matched}. Due to high uncertainty in regions without well data, the geological realisations have not been completely similar to the reference model, so the uncertainty of geology could be considered.

\begin{figure}
    \centering
    \includegraphics[width=0.75\textwidth]{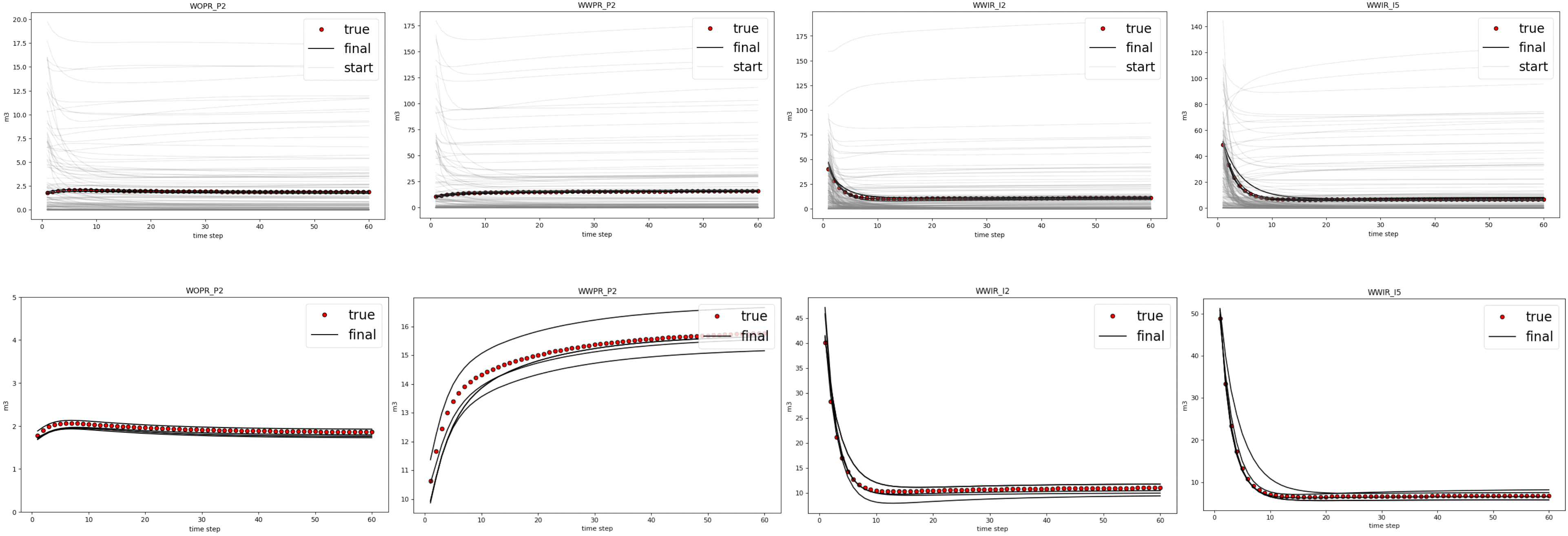}
    \caption{Production dynamics of injectors and producers for the initial model ensemble (top row) in comparison to the final optimised ensemble (bottom row) vs the reference (true) dynamic response} 
    \label{fig:fluvial_Rates_matched}
\end{figure}

In Figure \ref{fig:fluvial_Rates_matched}, oil and water production profiles for well P2 and injection profiles for wells I3 and I5 are presented. In grey lines there are dynamic responses of the initial population of generated models, which shows great variability due to the randomness of the initial population to spread across the LS evenly. Final optimised realisations are shown in solid black lines, in proximity to history data. Only three wells penetrated the channel, so the rest were performed without production. Values of production comparing reference and optimised models are quite close to each other, so the similarity in the dynamic response can be considered acceptable.

\begin{figure}
    \centering
    \includegraphics[width=0.75\textwidth]{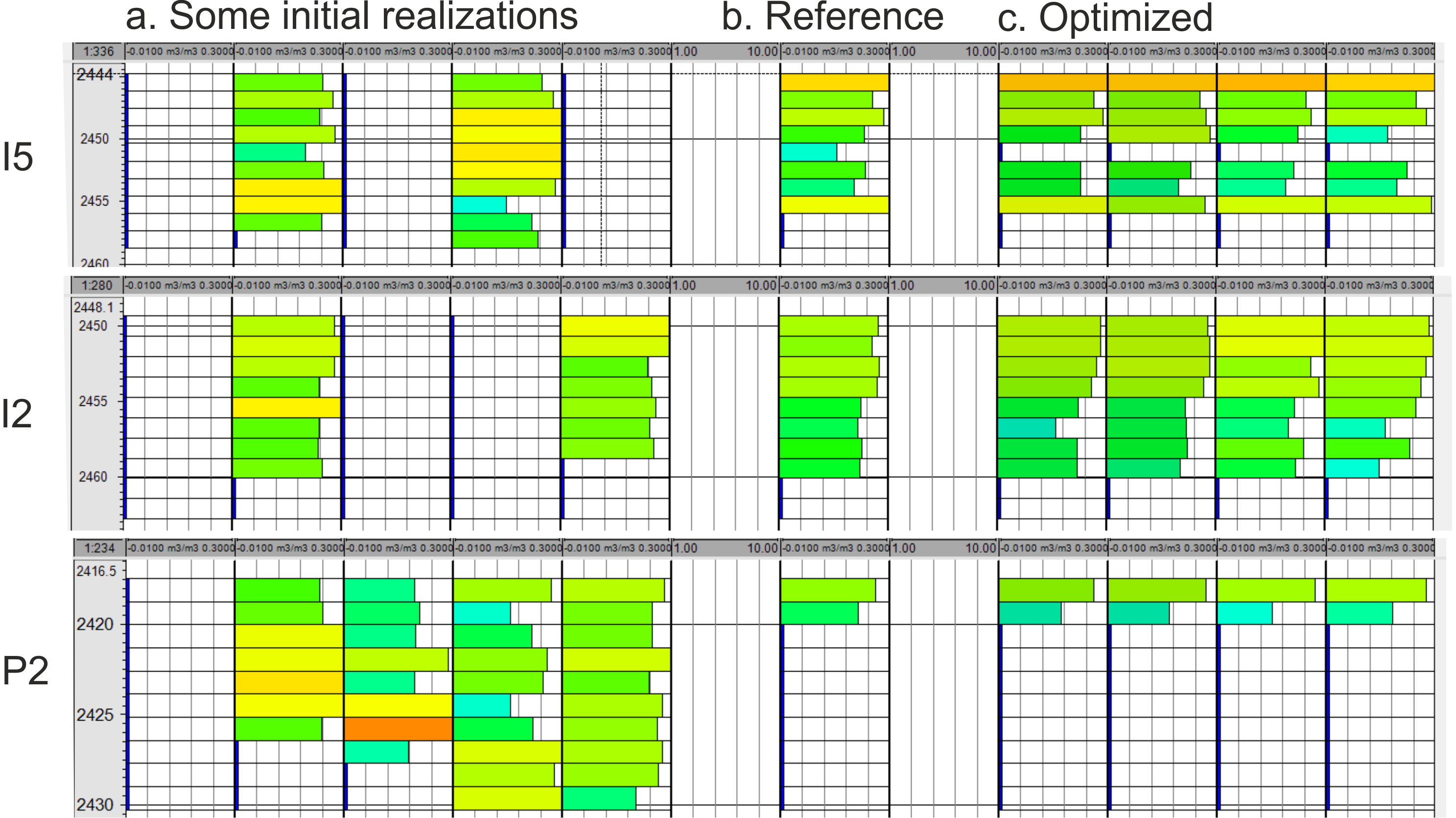}
    \caption{Static (porosity) property in wells P2, I2, I5: a. examples of initial (start) realisations, b. reference (true) realisation, c. optimised (final) realisations}
    \label{fig:fluvial_Crossections}
\end{figure}

Earlier, we considered the static well properties known to us, so the optimisation process should have found solutions similar to the reference model. GWAE along with the optimisation process, managed to find models with appropriate static properties, Figure \ref{fig:fluvial_Crossections}. It can be stated that the GWAE can be used to generate an ensemble of reliable realisations conditioned to well data.

\subsection{Ablation study}
To test the effectiveness of the OF as a tool for controlling the realism of GWAE generated realisations, the following experiment was conducted: The Realisation Loss component of the OF, which implicitly measures the realism of GWAE generated realisation, was disabled, then an optimisation process was run. Thus, we simulate a HM process in which the optimiser does not receive information about the spatial location of the geological representations in the LS, and the direction of the optimisation with respect to the location in the LS. 
As a result of HM, we expect the optimiser to find a geological models that provides similar static and dynamic components. However, the geological realism of the model is not expected, as no information has been obtained by the optimiser.

\begin{figure}
    \centering
    \includegraphics[width=0.75\textwidth]{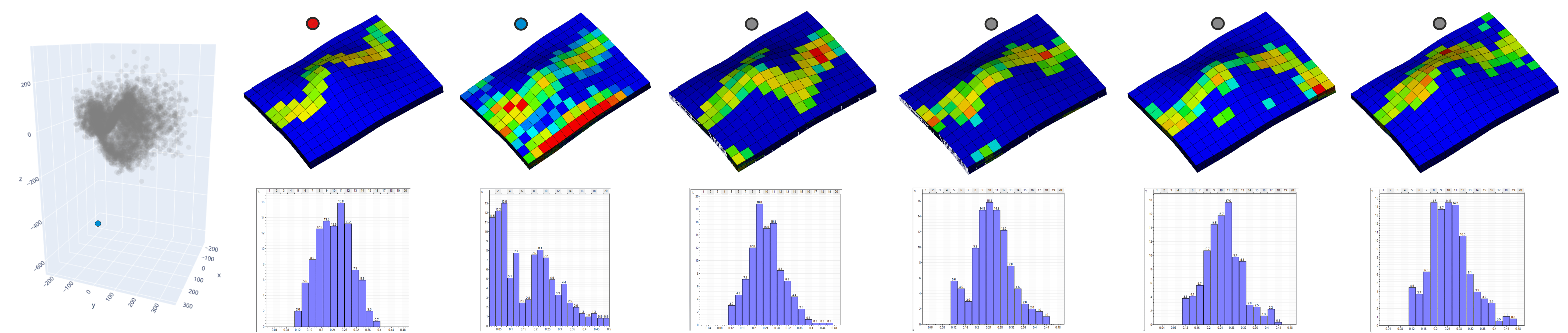}
    \caption{Left: spatial location of the geological model (blue dot) in the PCA representation. Right: the reconstructed geological model (blue dot) differs from the reference (red dot) and models with Realism Loss component (grey dots).}
    \label{fig:ablation study}
\end{figure}

The left part of Figure \ref{fig:ablation study} shows the spatial location of the geological representation found in the PCA space. It can be seen that the found location (blue dot) is significantly distant from the main group of points (grey dots - geologically consistent prior manifold). This suggests that the reconstructed geological realisation will be significantly different from the prior set, which is a violation of geological realism. The right part of the Figure confirms this, as the reconstructed geological model differs from the reference model (red dot) and the reconstructed realisations with realism control (grey dots).

The following Figures \ref{fig:ablation logs}, \ref{fig:Ablation_rates} show that the optimisation process was able to find implementations that provide reliable static and dynamic characteristics of geological realisation. 

\begin{figure}
    \centering
    \includegraphics[width=0.75\textwidth]{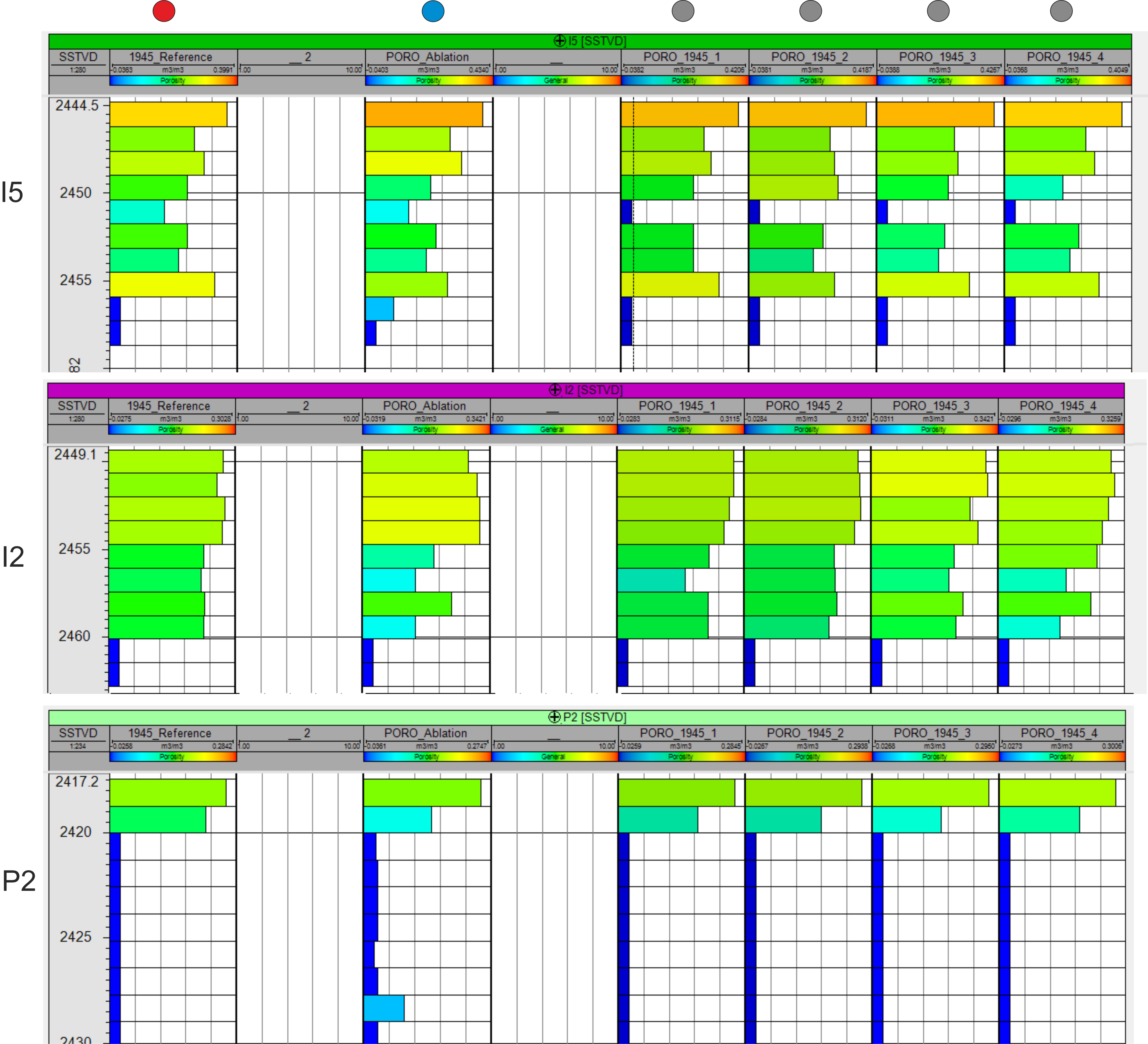}
    \caption{Static (porosity) property in wells P2, I2, I5. Reference model 
    (red dot), no- (blue dot) and with (grey dots) Realism Loss component. The overall similarity of well logs can be seen}     
    \label{fig:ablation logs}
\end{figure}

\begin{figure}
    \centering
    \includegraphics[width=0.75\textwidth]{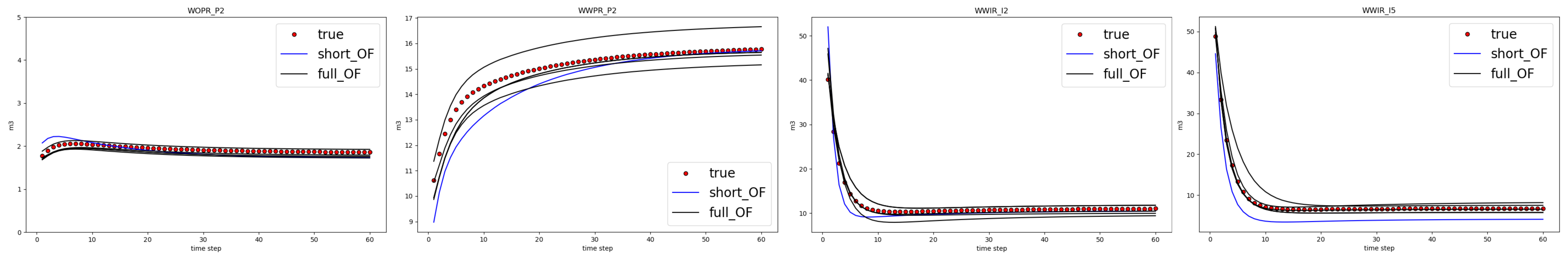}
    \caption{Production dynamics of injectors (I2, I5) and producer (P2) for the reference, no-Realism and with Realism Loss}
    \label{fig:Ablation_rates}
\end{figure}
\section{Conclusions}
We have shown the application of graph variational autoencoders to the problem of HM with geological realism control. It is demonstrated that graph architecture is a suitable approach for modelling geological structures. The use of a VAE architecture can effectively reduce the dimensionality of the original geological representation for search in LS. The LS itself has an internal structure that allows for a more efficient search for geological representations conditioned on static and dynamic well characteristics. The Geodesic metric is more suitable than the standard Euclidean for dealing with nonlinear hidden representations. A key aspect of optimising within the LS is incorporating an implicit control mechanism to ensure geological realism.
The key findings of the paper: 
\begin{itemize}
    \item GWAE approach to AHM in LS is suitable to account for multiple interpretation scenarios and non-uniqueness in model parameterisation;
    \item GWAE inherently represents reservoir topology rather than lattice-based deep learning networks;
    \item AHM with multicomponent OF that accounts and balances static/dynamic match and geological consistency via LS;
    \item LS continuity provides a way to navigate the search for HM models using evolutionary optimiser.
\end{itemize}

\section{Acknowledgment}
The authors appreciate the support of Rock Flow Dynamics for GeoDataScience \& UQ group research at Heriot-Watt University and the provision of tNavigator software licenses for reservoir flow simulations.
The authors are grateful to Gleb Shishaev's PhD examiners, Prof. Ahmed ElSheikh of Heriot-Watt University and Prof. Denis
Voskov of TU Delft for their careful review of the thesis and for providing valuable comments and advice that helped
improve this preprint.

\end{document}